\documentclass[prl,aps,twocolumn,showpacs,bibnotes,superscriptaddress,epsf]{revtex4}

\usepackage{graphicx}
\usepackage{dcolumn}
\usepackage{bm}
\usepackage{SIunits}
\usepackage{tabularx}

\begin{document}

\title{Breakdown of single spin-fluid model in heavily hole-doped superconductor CsFe$_2$As$_2$}
\author{D. Zhao}\affiliation{Hefei National Laboratory for Physical Science at Microscale and Department of Physics, University of Science and Technology of China, Hefei, Anhui 230026, People's Republic of China}
\author{S. J. Li}\affiliation{Hefei National Laboratory for Physical Science at Microscale and Department of Physics, University of Science and Technology of China, Hefei, Anhui 230026, People's Republic of China}
\author{N. Z. Wang}\affiliation{Hefei National Laboratory for Physical Science at Microscale and Department of Physics, University of Science and Technology of China, Hefei, Anhui 230026, People's Republic of China}
\author{J. Li}\affiliation{Hefei National Laboratory for Physical Science at Microscale and Department of Physics, University of Science and Technology of China, Hefei, Anhui 230026, People's Republic of China}
\author{D. W. Song}\affiliation{Hefei National Laboratory for Physical Science at Microscale and Department of Physics, University of Science and Technology of China, Hefei, Anhui 230026, People's Republic of China}
\author{L. X. Zheng}\affiliation{Hefei National Laboratory for Physical Science at Microscale and Department of Physics, University of Science and Technology of China, Hefei, Anhui 230026, People's Republic of China}
\author{L. P. Nie}\affiliation{Hefei National Laboratory for Physical Science at Microscale and Department of Physics, University of Science and Technology of China, Hefei, Anhui 230026, People's Republic of China}
\author{X. G. Luo}\affiliation{Hefei National Laboratory for Physical Science at Microscale and Department of Physics, University of Science and Technology of China, Hefei, Anhui 230026, People's Republic of China}\affiliation{Key Laboratory of Strongly-coupled Quantum Matter Physics, University of Science and Technology of China, Chinese Academy of Sciences, Hefei 230026, China}\affiliation{Collaborative Innovation Center of Advanced Microstructures, Nanjing University, Nanjing 210093, China}
\author{T. Wu}\email{wutao@ustc.edu.cn}\affiliation{Hefei National Laboratory for Physical Science at Microscale and Department of Physics, University of Science and Technology of China, Hefei, Anhui 230026, People's Republic of China}\affiliation{Key Laboratory of Strongly-coupled Quantum Matter Physics, University of Science and Technology of China, Chinese Academy of Sciences, Hefei 230026, China}\affiliation{Collaborative Innovation Center of Advanced Microstructures, Nanjing University, Nanjing 210093, China}
\author{X. H. Chen}\affiliation{Hefei National Laboratory for Physical Science at Microscale and Department of Physics, University of Science and Technology of China, Hefei, Anhui 230026, People's Republic of China}\affiliation{Key Laboratory of Strongly-coupled Quantum Matter Physics, University of Science and Technology of China, Chinese Academy of Sciences, Hefei 230026, China}\affiliation{Collaborative Innovation Center of Advanced Microstructures, Nanjing University, Nanjing 210093, China}

\begin{abstract}
Although Fe-based superconductors are multiorbital correlated
electronic systems, previous nuclei magnetic resonance (NMR)
measurement suggests that a single spin-fluid model is sufficient to
describe its spin behavior. Here, we firstly observed the breakdown
of single spin-fluid model in a heavily hole-doped Fe-based
superconductor CsFe$_2$As$_2$ by site-selective NMR measurement. At
high temperature regime, both of Knight shift and nuclei
spin-lattice relaxation at $^{133}$Cs and $^{75}$As nuclei exhibit
distinct temperature-dependent behavior, suggesting the breakdown of
single spin-fluid model in CsFe$_2$As$_2$. This is ascribed to the
coexistence of both localized and itinerant spin degree of freedom
at 3\emph{d} orbits, which is consistent with orbital-selective Mott
phase. However, single spin-fluid behavior is gradually recovered by
developing a coherent state among 3\emph{d} orbits with decreasing
temperature. A Kondo liquid scenario is proposed for the
low-temperature coherent state. The present work sets strong
constraint on the theoretical model for Fe-based superconductors.

\end{abstract}

\pacs{74.70.Xa, 74.25.nj, 71.27.+a}

\maketitle

In high-T$_c$ cuprate superconductors, single spin-fluid model has
been widely adopted as a theoretical starting point although at
least one 3\emph{d} and two 2\emph{p} orbits from copper and oxygen
sites should be considered together in theoretical model in
principle\cite{Lee}. The basis of such hypothesis is mostly based on
the celebrated concept of Zhang-Rice singlet\cite{Zhang}, which
successfully converts the complex reality into single band t-J
model. Such single spin-fluid model has been validated by early
site-selective nuclei magnetic resonance (NMR) measurement on
$^{89}$Y, $^{63}$Cu and $^{17}$O nuclei in
YBa$_2$Cu$_3$O$_{6+x}$\cite{Alloul, Takigawa}.

In Fe-based superconductors, the multiorbital nature is a key factor
to understand its basic properties\cite{Yi}. Considering correlation
effect due to Hund$^\prime$s coupling, single spin-fluid model
should be insufficient in this case\cite{Yin2}. However, previous
site-selective NMR measurement on F-doped LaOFeAs found that Knight
shift and nuclei spin-lattice relaxation on different nuclei are
nearly identical\cite{Grafe}, including $^{75}$As, $^{57}$Fe,
$^{19}$F and $^{139}$La nuclei. This result suggests a single
spin-fluid model, which is consistent with weak coupling theory
based on itinerant nature of Fe 3\emph{d} electrons\cite{Hirchfeld,
Chubukov}. Similar behavior was also observed in many other Fe-based
superconductors\cite{Ma1, Kitagawa, Ma2}. On the other hand, strong
coupling theory based on local nature of Fe 3\emph{d} electrons has
also been proposed for Fe-based superconductors\cite{Yu}, in which
the coexistence of itinerant and localized electrons at different
3\emph{d} orbits would appear in a so-called orbital-selective Mott
phase\cite{Anisimov, Medici1, Georges}. Recently, orbital-selective
Mott phase has been widely observed in FeSe-derived superconductors
by angle-resolved photoemission spectroscopy (ARPES)\cite{Yi1, Yi2}.
However, site-selective NMR experiment has not yet observed any
breakdown of single spin-fluid model in FeSe-derived
superconductors\cite{Ma1}.

Very recently, a similar orbital-selective Mott phase has also been
proposed in heavily hole-doped Fe-based superconductors
AFe$_2$As$_2$ (A = K, Rb, Cs)\cite{Hardy, Medici2}. Furthermore, a
so-called ``Knight shift anomaly'' phenomenon has been observed by
$^{75}$As NMR in AFe$_2$As$_2$ (A = K, Rb, Cs)\cite{Wu}, which hints
a possible deviation from single spin-fluid model. In order to
further examine the single spin-fluid model in heavily hole-doped
Fe-based superconductors AFe$_2$As$_2$ (A = K, Rb, Cs), we conducted
a site-selective NMR experiment on CsFe$_2$As$_2$ single crystal by
measuring $^{133}$Cs and $^{75}$As nuclei.

\begin{figure}[t]
\centering
\includegraphics[width=0.45\textwidth]{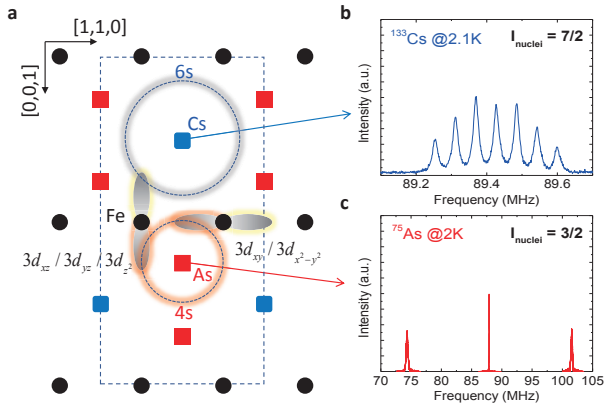}
\caption{(color online). (a) Illustration of the microscopic process
of the transferred hyperfine interaction on $^{75}$As and $^{133}$Cs
nuclei. The crystalline structure of CsFe$_2$As$_2$ is shown in the
side view along Fe-Fe direction. The hyperfine interaction of
$^{75}$As and $^{133}$Cs are dominated by the transferred hyperfine
interaction through the hybridization of onsite 4\emph{s} or
6\emph{s} orbit with 3\emph{d} orbits at the nearest-neighbour Fe
atoms. We performed NMR measurements on both $^{75}$As and
$^{133}$Cs nuclei. The full spectrum at 2 K are shown in (b) for
$^{133}$Cs nuclei and (c) for $^{75}$As nuclei.}
\end{figure}

As shown in Fig.1a, considering the spatial distribution of
3\emph{d} orbits, the 3\emph{d} orbits at Fe sites could be divided
into two categories. The first category includes 3$d_{xy}$ and
3$d_{x^2+y^2}$ and the second one includes 3$d_{xz}$, 3$d_{yz}$ and
3$d_{z^2}$. It's very obvious that the orbits in second category
have much larger spatial distribution along c-axis than that in the
first one. In Fe-based superconductors, the dominated hyperfine
interaction between 3\emph{d} electrons at Fe site and $^{75}$As
nuclei is verified to be transferred hyperfine interaction mediated
by 4\emph{s} orbit of $^{75}$As\cite{Kitagawa2}. In this case, the
overlap between 3\emph{d} and 4\emph{s} orbits would largely
determine the strength of the transferred hyperfine interaction.
Therefore, in principle, the electrons on different 3\emph{d} orbits
have different transferred hyperfine interaction with $^{75}$As
nuclei as shown in Fig.1a. We could express the spin part of Knight
shift from different 3\emph{d} orbits as the following equation:
\begin{equation}\label{1}
K_{s}(T) = \sum_{\sigma} A_{\sigma}\chi_{\sigma}(T), (\sigma = xz, yz, xy, z^{2}, x^{2}+y^{2})
\end{equation}
$A_{\sigma}$ is the hyperfine coupling tensor from different
3\emph{d} orbits. $\chi_{\sigma}(T)$ is the orbital-dependent local
spin susceptibility. Considering similar transferred hyperfine
interaction, the $^{133}$Cs nuclei in spacer layer with distinct
out-of-plane distance to Fe site would only have significant overlap
with the 3\emph{d} orbits in above second category, suggesting an
orbital-selective NMR probe. Based on this fact, the site-selective
NMR by measuring both $^{75}$As and $^{133}$Cs nuclei would have the
ability to examine single spin-fluid model in our case. Our
following results unambiguously confirm the breakdown of single
spin-fluid model in CsFe$_2$As$_2$.

High-quality CsFe$_2$As$_2$ single crystals are grown by the
self-flux technique\cite{Wang}. All NMR measurement on $^{75}$As and
$^{133}$Cs nuclei are conducted from 2 K to 300 K under an external
magnetic field of 14 Tesla parallel to c axis. The nuclei spin
number $I_{nuclei}$ for $^{75}$As and $^{133}$Cs nuclei are 3/2 and
7/2 respectively. The standard full spectrum of $^{75}$As and
$^{133}$Cs nuclei are shown in Fig.1b. There are three transition
lines for $^{75}$As nuclei and seven transition lines for $^{133}$Cs
nuclei. For $^{133}$Cs nuclei, all NMR peaks have a similar
linewidth of $\sim$20 KHz at 2 K, suggesting a magnetic broadening
origin. This is also consistent with small quadrupole frequency
below. For $^{75}$As nuclei, the linewidth for satellite peaks and
central peak are $\sim$300 KHz and $\sim$30 KHz at 2 K, indicating
that the linewidth of satellite peaks is dominated by quadrupole
broadening. Similar behavior has already been seen in previous
study\cite{Li}. By measuring the separation between each transition
lines, the quadrupole frequency $\upsilon$$_Q$ for $^{75}$As and
$^{133}$Cs nuclei are determined to be $\sim$13.6 MHz and
$\sim$0.058 MHz respectively. Both of Knight shifts for $^{75}$As
and $^{133}$Cs nuclei are determined by measuring the frequency
position of the central transition line. Nuclei spin-lattice
relaxations rate 1/T$_1$ are also measured on the central transition
line for both $^{75}$As and $^{133}$Cs nuclei.

The main results in this letter are shown in Fig.2. As shown in
Fig.2a, the temperature-dependent Knight shift of $^{75}$As
($^{75}$\emph{K}) nuclei exhibits a characteristic crossover
behavior. At high temperature regime, $^{75}$\emph{K} is gradually
increasing as temperature decreasing. This is also consistent with
high-temperature bulk magnetic susceptibility, which suggests a
localized spin behavior\cite{Wu}. As temperature further decreasing,
$^{75}$\emph{K} shows a maximum and then starts to decrease with
lowering temperature. Below 20 K, $^{75}$\emph{K} becomes saturated
and shows a temperature-independent behavior. Above
temperature-dependent behavior has been ascribed to an
incoherent-to-coherent electronic crossover, which is also observed
in KFe$_2$As$_2$ and RbFe$_2$As$_2$ with different crossover
temperature\cite{Wu}. In sharp contrast, the remarkable crossover
behavior in temperature-dependent Knight shift of $^{75}$As nuclei
is completely absent in that of $^{133}$Cs nuclei. As shown in
Fig.2a, the temperature dependent Knight shift of $^{133}$Cs
($^{133}$\emph{K}) nuclei shows a monotonous decreasing in the whole
temperature range. The localized spin behavior as shown in
high-temperature bulk susceptibility and $^{75}$\emph{K} does not
show up in $^{133}$\emph{K}. This result indicates that the
localized spin behavior probably comes from 3$d_{xy}$ or
3$d_{x^2+y^2}$ orbits. As we mentioned before, the hyperfine
interaction between 3\emph{d} electrons and $^{133}$Cs nuclei is
dominated by the transferred hyperfine interaction which is
dependent on the overlap between 3\emph{d} orbits at Fe site and
6\emph{s} orbits at Cs site. Since both of 3$d_{xy}$ or
3$d_{x^2+y^2}$ orbits at Fe site has less overlap with 6\emph{s}
orbits at Cs site due to limited spatial distribution along c axis,
the transferred hyperfine interaction would be not sensitive to
3$d_{xy}$ or 3$d_{x^2+y^2}$ orbits. In this case, we could ascribe
the origin of electronic crossover behavior observed by $^{75}$As
nuclei to the 3$d_{xy}$ or 3$d_{x^2+y^2}$ orbits. Based on previous
theoretical calculation\cite{Hardy, Backes}, the 3$d_{x^2+y^2}$
orbit has much less mass renormalization effect than other 3\emph{d}
orbits. Considering this point, the high-temperature localized spin
behavior could only be ascribed to 3$d_{xy}$ orbit, which is also
consistent with that the 3$d_{xy}$ orbit has the maximum mass
renormalization effect among all 3\emph{d} orbits\cite{Hardy,
Backes}. Therefore, we believe that the temperature dependence of
$^{133}$\emph{K} is dominated by itinerant 3\emph{d} orbits but the
temperature dependence of $^{75}$\emph{K}, especially for the
incoherent-to-coherent crossover, is mainly dominated by localized
3$d_{xy}$ orbit. This result strongly proves the breakdown of single
spin-fluid model in CsFe$_2$As$_2$. By scaling both
temperature-dependent $^{75}$\emph{K} and $^{133}$\emph{K}, we found
an identical temperature dependent behavior below T$^\ast$ $\sim$ 75
K. As shown in the inset of Fig.2a, single spin-fluid behavior is
recovered below T$^\ast$. This result indicates that, although
single spin-fluid model is broken above T$^\ast$ due to coexistence
of localized and itinerant 3\emph{d} electrons, a coherent state
involving both localized and itinerant 3\emph{d} electrons appears
below T$^\ast$ which still follows single spin-fluid model. Similar
crossover behavior was also observed in FeSe-derived superconductors
by ARPES, in which the high-temperature incoherent state is ascribed
to orbital-selective Mott phase\cite{Yi1, Yi2}. Here, we also
believe that a similar orbital-selective Mott phase might appear
above T$^\ast$ in CsFe$_2$As$_2$. Low-temperature ARPES and STM
experiments have already observed a coherent peak from
3\emph{d}$_{xy}$ orbit close to Fermi level below 20 K in
KFe$_2$As$_2$\cite{Fang}. Further ARPES experiment with whole
temperature range is needed to verify the exact nature of the
high-temperature incoherent state in CsFe$_2$As$_2$.

\begin{figure}[t]
\centering
\includegraphics[width=0.45\textwidth]{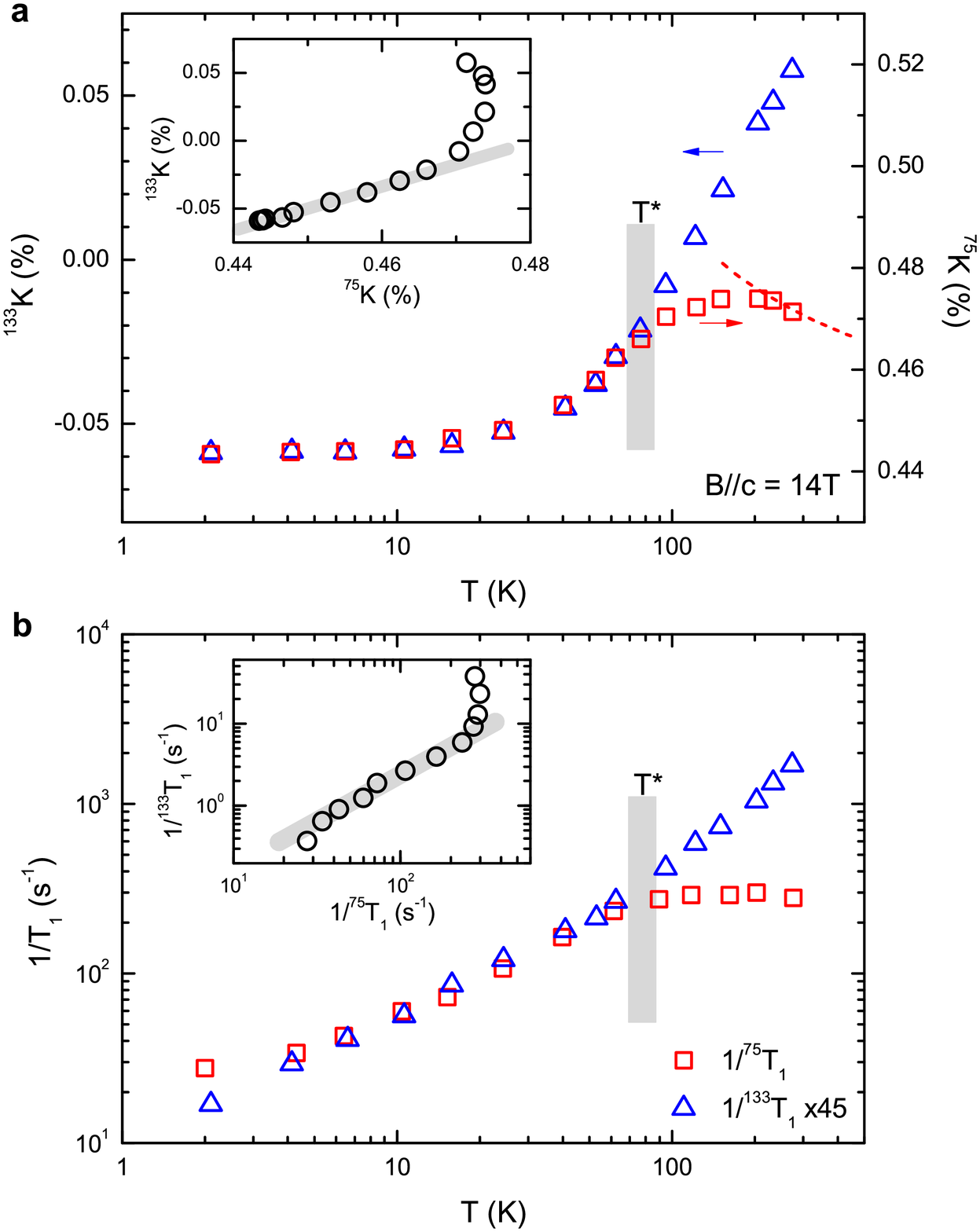}
\caption{(color online). (a) Temperature dependent Knight shift for
both $^{75}$As and $^{133}$Cs nuclei. The blue triangles represent
the Knight shift of $^{133}$Cs nuclei. The red squares represents
the Knight shift of $^{75}$As. The red dash line is a guiding line
for temperature dependent $^{75}$\emph{K} at high temperature. Both
Knight shifts below T$^\ast$ follow a same temperature dependent
behavior, which is also confirmed by the $^{133}$\emph{K} -
$^{75}$\emph{K} plot in the inset. (b)Temperature dependent nuclei
spin-lattice relaxation for both $^{75}$As and $^{133}$Cs nuclei.
The blue triangles represent the nuclei spin-lattice relaxation of
$^{133}$Cs nuclei. The red squares represents the nuclei
spin-lattice relaxation of $^{75}$As. Both nuclei spin-lattice
relaxation below T$^\ast$ follow a same power-law behavior, which is
also confirmed by the $\frac{1}{^{133}T_1}$ - $\frac{1}{^{75}T_1}$
plot in the inset. The fitting formula of spin-lattice relaxation
decay is
$I(t)=I_0+I_1(0.1e^{-(\frac{t}{T_1})^r}+0.9e^{-(\frac{6t}{T_1})^r})$
for $^{75}$As nuclei and $I(t)=I_0+I_1e^{-(\frac{t}{T_1})^r}$ for
$^{133}$Cs nuclei.}
\end{figure}

In Fig.2b, we also measured the temperature dependent 1/T$_1$ for
both $^{75}$As and $^{133}$Cs nuclei. Similar to Knight shift
result, the temperature-dependent 1/T$_1$ also shows a distinct
temperature-dependent behavior for both $^{75}$As and $^{133}$Cs
nuclei above T$^\ast$. For $^{133}$Cs nuclei, the
temperature-dependent 1/T$_1$ follows an approximate power-law
behavior with 1/T$_1$ $\sim$ T$^{1.36}$, which looks like a strange
metallic behavior. For $^{75}$As nuclei, the temperature-dependent
1/T$_1$ is almost temperature independent, which is consistent with
a localized moment behavior\cite{Wu}. This result further confirms
the above conclusion on the breakdown of single spin-fluid model
above T$^\ast$. Below T$^\ast$, an identical temperature dependent
behavior for both $^{75}$As and $^{133}$Cs nuclei with 1/T$_1$
$\sim$ T$^{0.75}$ appears as shown in the inset of Fig.2b. This
result is consistent with the formation of a coherent state below
T$^\ast$. A deviation from T$^{0.75}$ power-law behavior is also
observed at very low temperature. This is due to the appearance of
two-component behavior in spin-lattice relaxation decay. The details
has been discussed in our previous work and we found that such
two-component behavior is dependent on external magnetic
field\cite{Wu, Li}. When we measure nuclei spin-lattice relaxation
under zero field, a perfect spin-lattice relaxation decay with
single component is observed and the T$^{0.75}$ power-law behavior
is extended to the lowest temperature\cite{Wu}. Based on this
observation, we ascribe the low-temperature T$^{0.75}$ power-law
behavior to the characteristic property of the low-temperature
coherent state. The field-induced deviation from T$^{0.75}$
power-law behavior due to two-component behavior in spin-lattice
relaxation decay might be other novel effect but this is not the
focus in this letter.

\begin{figure}[t]
\centering
\includegraphics[width=0.45\textwidth]{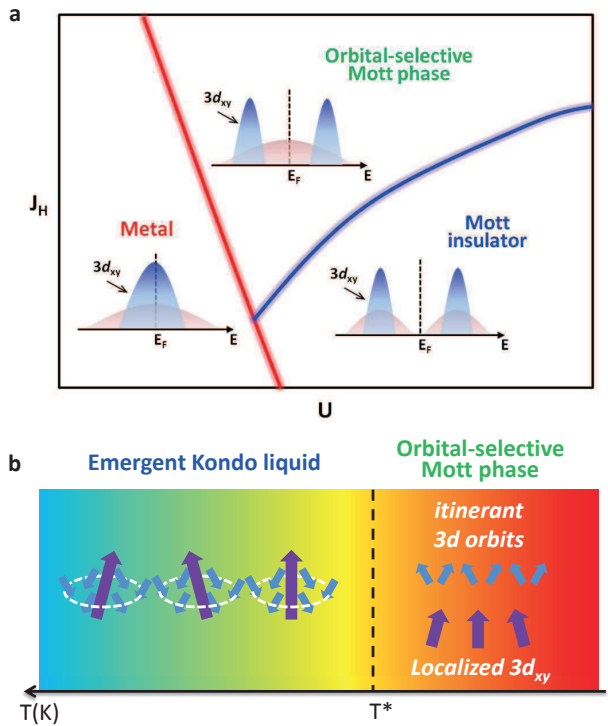}
\caption{(color online). (a) Schematic phase diagram of Fe-based
superconductors tuned by on-site Coulomb repulsion U and
Hund$^\prime$s coupling J$_H$\cite{Medici1, Georges}. When both of U
and J$_H$ are small, the system is a normal metal. When J$_H$ is
moderate but U is enough large, the system will enter into Mott
insulating phase. When J$_H$ become enough large, an
orbital-selective Mott phase will replace the Mott insulating phase
in phase diagram. (b) Illustration of microscopic picture for the
incoherent-to-coherent crossover. At high temperature, the system
behaves as orbital-selective Mott phase with both localized and
itinerant 3\emph{d} electrons. Below T$^\ast$, the system enters
into a Kondo liquid state, in which all 3\emph{d} electron becomes
coherent through Kondo-type coupling.}
\end{figure}

Based on above results on Knight shift and nuclei spin-lattice
relaxation, we found that the high-temperature incoherent state
above T$^\ast$ looks like a so-called orbital-selective Mott
phase\cite{Anisimov, Medici1, Georges}, in which 3\emph{d$_{xy}$}
orbit is probably localized and other 3\emph{d} orbits remain
itinerant. As shown in Fig.3a, orbital-selective Mott phase has
already been proposed in phase diagram based on strong coupling
theory\cite{Yu, Medici1}. Previous theoretical study suggests that
the 3\emph{d$_{xy}$} orbit would firstly occurs orbital-selective
Mott localization among all 3\emph{d} orbits in AFe$_2$As$_2$ (A =
K, Rb, Cs)\cite{Hardy, Backes}. This is consistent with our present
site-selective NMR results. In detail, the contrasting temperature
dependence of $^{133}$\emph{K} and $^{75}$\emph{K} suggests that the
orbital-selective Mott localization or at least the strongest
correlation effect happens on 3\emph{d$_{xy}$} orbit, which leads to
a localized spin behavior above T$^\ast$. Further ARPES experiment
would be expected to verify the exact nature of 3\emph{d$_{xy}$}
orbit above T$^\ast$.

Below T$^\ast$, a coherent state between localized and itinerant
3\emph{d} orbits is developed, which is also verified by
low-temperature ARPES and STM results on KFe$_2$As$_2$\cite{Fang}.
How to understand the underlying mechanism of such
incoherent-to-coherent crossover? In fact, incoherent-to-coherent
crossover has already been proposed in early dynamical mean field
theory (DMFT) by K. Haule and G. Kotliar\cite{Haule}, in which each
3\emph{d} orbit would have a different crossover temperature and the
lowest one determines the crossover temperature for whole
system\cite{Yin}. Recently, this scenario has been further developed
by using numerical renormalization group as a viable multi-band
impurity solver for DMFT, suggesting that strong Kondo-type
screening correlation exists during the incoherent-to-coherent
crossover\cite{Stadler}. On the other hand, our previous NMR study
also observed ``Knight shift anomaly'' and relevant scaling behavior
in AFe$_2$As$_2$ (A = K, Rb, Cs), which are ascribed to emergent
Kondo lattice behavior of 3\emph{d} electron system\cite{Wu}. Based
on all these facts, as shown in Fig.3b, we proposed that the
emergent coherent state below T$^\ast$ might be treated as a Kondo
liquid state similar as that in heavy fermion system with
\emph{f}-electron\cite{Yang1, Yang2}.

In \emph{f}-electron heavy fermion system, Kondo liquid state is an
emergent state due to collective Kondo coupling between localized
and itinerant electrons. In Kondo liquid state, the localized spin
degree of freedom would be screened out by itinerant electrons,
which leads to the decomfinement of localized moments. In our case,
we also believe that a similar decomfinement of localized 3\emph{d}
electrons happens due to certain Kondo-type coupling between
localized and itinerant 3\emph{d} electrons, such as off-site Kondo
coupling\cite{Anisimov2}. If this scenario is finally validated, the
Kondo-type coupling between localized and itinerant spin degree of
freedom for 3\emph{d} electrons should be an important ingredient in
effective theoretical model for Fe-based superconductors\cite{Weng}.
This would stimulate further theoretical investigation in
AFe$_2$As$_2$ (A = K, Rb, Cs) system and bring new understanding on
the mechanism of superconducting pairing in Fe-based
superconductors.

Finally, we would like to address that the strong coupling feature
observed in AFe$_2$As$_2$ (A = K, Rb, Cs) is probably due to the
Mott insulating phase at 3\emph{d}$^5$ configuration\cite{Yu,
Medici2}. Based on DMFT calculation, the realization of Mott
insulating phase would be much easier at half filling than other
integer fillings\cite{Georges}. For example, the critical mutual
Coulomb replusion (U$_c$) for Mott transition has a minimum at half
filling. For 3\emph{d} electron system, the half filling is
3\emph{d}$^5$ configuration. In Fe-based superconductors, the parent
compound with 3\emph{d}$^6$ configuration is indeed a bad metal but
not Mott insulator. However, we could expect a real Mott insulating
phase appears at 3\emph{d}$^5$ configuration and this has already
been proposed in previous theory\cite{Medici2}. In this case, when
we doped considerable holes from parent compound with 3\emph{d}$^6$
configuration, the system is actually approaching to Mott insulating
phase at 3\emph{d}$^5$ configuration. Therefore, the emergence of
orbital-selective Mott phase between 3\emph{d}$^5$ configuration and
3\emph{d}$^6$ configuration is not a surprising in strong coupling
scenario. In sharp contrast, when we doped electrons from parent
compound with 3\emph{d}$^6$ configuration, the correlation effect
would becomes weaker during the evolution from 3\emph{d}$^6$
configuration towards 3\emph{d}$^7$ configuration\cite{Yu, Georges}.
This is consistent with experimental results on Co-doped
BaFe$_2$As$_2$\cite{Medici2}. In cuprate superconductors, the most
novel strongly correlated physics (such as pesudogap, stripe,
nematicity, charge order, etc.) always emerges in the underdoped
regime with hole doping\cite{Keimer}. Here, in Fe-based
superconductors, considering above Mott physics at 3\emph{d}$^5$,
the most novel strongly correlated physics would happen in the
overdoped regime with hole doping from parent compound at
3\emph{d}$^6$. It needs further experimental and theoretical survey
to figure out the exact nature of strongly correlated physics in
this regime.

This work is supported by the National Key R$\&$D Program of the
MOST of China (Grant No. 2016YFA0300201£¬2017YFA0303000), the National Natural
Science Foundation of China (Grants No. 11522434, 11374281,
U1532145), Science Challenge Project (Grants No. TZ2016004), the
Fundamental Research Funds for the Central Universities and the
Chinese Academy of Sciences. T. W. acknowledges the Recruitment
Program of Global Experts and the CAS Hundred Talent Program.

\end{document}